\documentclass[trackchanges, onecolumn]{aastex631}

\usepackage{wrapfig}
\usepackage{rotating}
\shorttitle{RS Oph Super-remnant}
\shortauthors{Shara, Lanzetta, Masegian et al.}

\begin{document}

%%
%\watermark{DRAFT}
%%%%%%%%%%%%%%%%%%%%%%%%%%%%%%%%%%%%%%%%%%

\title[RS OPH nova Super-remnant]{A 70 pc-Diameter Nova Super-remnant Surrounding the Recurrent Nova RS Ophiuchi}

\correspondingauthor{Michael Shara  mshara@amnh.org}

\author[0000-0003-0155-2539]{Michael M. Shara}
\affiliation{Department of Astrophysics, American Museum of Natural History, New York, NY 10024, USA}

\author[0000-0001-6906-7594]{Kenneth M. Lanzetta}
\affiliation{Department of Physics and Astronomy, Stony Brook University, Stony Brook, NY 11794-3800, USA}

\author[0000-0002-3361-2893]{Alexandra Masegian}
\affiliation{Department of Astronomy, Columbia University, 538 West 120th Street, New York, NY 10027, USA}

\author[0000-0003-2922-1416]{James T. Garland}
\affiliation{Department of Astrophysics, American Museum of Natural History, New York, NY 10024, USA}
\affil{David Dunlap Department of Astronomy, University of Toronto, 50 George Street, Toronto, ON M5S 3H4, Canada}

\author[0009-0006-3617-1356]{Stefan Gromoll}
\affiliation{Amazon Web Services, 410 Terry Ave. N, Seattle, WA 98109, USA}

\author[0000-0001-8646-0419]{Alexei Kniazev}
\affiliation{South African Astronomical Observatory, Observatory Road, Observatory 7925, Capetown, South Africa}
\affil{Southern African Large Telescope, PO Box 9, Observatory 7935, South Africa}
\affil{Sternberg Astronomical Institute, Lomonosov Moscow State University, Moscow, Russia}

\author[0000-0001-9788-3345]{Lee Townsend}
\affil{South African Astronomical Observatory, Observatory Road, Observatory 7925, Capetown, South Africa}
\affil{Southern African Large Telescope, PO Box 9, Observatory 7935, South Africa}

\author[0009-0008-4599-2935]{David Zurek}
\affiliation{Department of Astrophysics, American Museum of Natural History, New York, NY 10024, USA}

\author[0000-0001-7796-1756]{Joanna Mikolajewska}
\affil{N. Copernicus Astronomical Center, Polish Academy of Sciences, Bartycka 18, 00–716 Warsaw, Poland}

\author[0000-0002-9821-2911]{David Valls-Gabaud}
\affil{Observatoire de Paris, LUX, CNRS UMR 8112, 61 Avenue de l’Observatoire, 75014 Paris, France}

\author[0000-0001-7796-1756]{Frederick M. Walter}
\affil{Department of Physics and Astronomy, Stony Brook University, Stony Brook, NY 11794-3800, USA}

\author[0000-0002-0004-9360]{John K. Webb}
\affil{Institute of Astronomy, University of Cambridge, Madingley Road, Cambridge CB3 0HA, United Kingdom}

\begin{abstract}
Recurrent novae undergo thermonuclear-powered eruptions separated by less than 100 years, enabled by subgiant or red giant donors transferring hydrogen-rich matter at very high rates onto their massive white dwarf companions. The most-rapidly moving parts of envelopes ejected in successive recurrent nova events are predicted to overtake and collide with the slowest ejecta of the previous eruption, leading to the buildup of vast ($\sim$ 10 - 100 parsec) super-remnants surrounding all recurrent novae; but only three examples are currently known. We report deep narrowband imaging and spectroscopy which has revealed a $\sim$ 70-parsec-diameter shell surrounding the frequently recurring nova RS Ophiuchi. We estimate the super-remnant mass to be $\sim$ 20 - 200 $M_{\odot}$, expanding at a few tens of km/s, with an age of order 50-100 kyr. Its extremely low surface brightness and large angular size help explain the hitherto surprising absence of nova super-remnants. Our results support the prediction that {\it all} recurrent novae are surrounded by similar extended structures.
\end{abstract}


\section{Introduction} \label{sec:intro}
\subsection{Novae}
Sufficient matter accreted into a white dwarf's envelope from its binary companion leads to a thermonuclear runaway \citep{starrfield1972,prialnik1979,Yaron2005} observed as a nova eruption. Reaching luminosities of $10^{4}$ to $10^{6}$$L_{\odot}$, nova eruptions continue for days to years, until most of the accreted envelope is ejected and nuclear reactions cease \citep{prialnik1978}. Novae undergo thousands of eruptions, usually separated by many thousands of years, during their multi-Gyr lifetimes \citep{Ford1978,Hillman2020}. 

\subsection{Recurrent Novae and Nova Super-remnants}
Recurrent novae (by definition) undergo eruptions separated by less than 100 years \citep{Schaefer2010}, enabled by subgiant or red giant donors transferring hydrogen-rich matter at very high rates (of order $10^{-7}M_{\odot}$/yr) onto a nearly Chandrasekhar-mass white dwarf \citep{Hillman2016}. The most-rapidly moving parts of envelopes ejected in successive recurrent nova events are predicted to overtake and collide with the slowest ejecta of the previous eruption, leading to the buildup of vast ($\sim$ 10 - 100 parsec) nova super-remnants (NSR) which should surround {\it all} recurrent novae \citep{Healy-Kalesh2023}. While much smaller ($<$ 1 parsec-sized) ejecta are known to surround the recurrent nova T Pyx \citep{Shara2015}, only two of the eleven known Galactic recurrent novae (KT Eri and T CrB) have identified NSR \citep{Shara2024a,Shara2024b}. The only example of an extragalactic NSR (over 100 pc in size!) is M31-12a in the Andromeda galaxy \citep{Darnley2019}. The most comprehensive search to date found no new examples of NSR surrounding 20 recurrent nova in the Local Group of galaxies \citep{Healy-Kalesh2024b}, where at least part of the reason for the lack of success is likely the extremely low surface brightness of NSR.   

\subsection{RS Oph}
The well-studied Galactic recurrent nova RS Oph \citep{Pottasch1967,Bode1985,Dobrzycka1996,Brandi2009,Mikolajewska2017} is located at a Gaia-determined distance of 2.68 $\pm$ 0.16 kpc \citep{schaefer2022}. Its Galactic latitude and longitude of +5.5 deg and +15.7 deg, respectively, place it $\sim$ 260 pc above the midplane of the Milky Way. The underlying binary, with an orbital period of 453.6 d, comprises a massive white dwarf which accretes hydrogen-rich matter from its red giant companion. RS Oph underwent nova eruptions in 1898, 1907, 1933, 1958, 1967, 1985, 2006 and 2021, reaching about fifth magnitude each time. RS Oph emits from the radio \citep{Hjellming1986} through the infrared \citep{Evans1988}, optical \citep{Blair1983}, ultraviolet \citep{Shore1996}, X-ray \citep{Sokoloski2006} and gamma ray \citep{HESS2022} portions of the electromagnetic spectrum, as well as extremely energetic protons \citep{Acciari2022}. Material ejected during these recurrent nova eruptions has been directly imaged and modeled \citep{OBRIEN2006,Bode2007,Booth2016}, extending a few arcseconds from the star. This pre-existing material, surrounding RS Oph, will inevitably be overtaken by photons emitted during future eruptions, which might lead to fluorescent and scattered-light light echoes. In addition, a degree-sized cavity surrounding RS Oph has recently been reported \citep{Healy-Kalesh2024a}, suggesting that a NSR may be present. So motivated, we have used the Condor Array Telescope to deeply image the environs of RS Oph, and obtained followup spectroscopy with SALT of the newly detected nebulosities.

\subsection{The Condor Telescope Array}
The Condor Array Telescope, located near Animas, New Mexico, is a grouping of six co-aligned 180 mm apochromatic refractors, each with its own low read-noise 9576 $\times$ 6388 pixel CMOS camera. It is designed to minimize background and scattered light in searches for low surface brightness objects \citep{Lanzetta2023}. Condor is equipped with a set of narrowband filters (each of either 3 or 4 nm FWHM, one per telescope): HeII (468.6~nm), [OIII] (500.7 nm), HeI (587.6 nm), H$\,\alpha$ (656.3 nm), [NII] (658.4 nm), and [SII] (671.6 nm). It is also equipped with a Luminance (L-band) filter with central wavelength = 550 nm, $>$ 97\% transmittance over the wavelength range 420-700 nm, and essentially zero transmittance outside that range. The very broad wavelength response of the L-band filter ($\sim$ 100X wider than that of the narrowband filters described above) ensures that line emission losses in difference images (narrowband minus L image) are negligible. 

Condor's demonstrated successes in finding extremely low surface brightness features surrounding the old nova (and dwarf nova) Z Cam \citep{shara2023}, the putative recurrent nova KT Eri \citep{Shara2024a} and the closest of all known recurrent novae, T CrB, \citep{Shara2024b} have prompted us to use it to search for other recurrent NSR, including that of RS Oph.

In Section 2 we describe the observational data of this paper. The images of the newly discovered nebulosity surrounding RS Oph are presented in Section 3, while their spectra are presented in Section 4. The implications of the images and spectra are discussed in Section 5. We summarize our results in Section 6.

\section{The Datasets}\label{sec:datasets}
\subsection{Condor Imaging}

The Condor field of view is 2.2 x 1.5 degree$^2$, with a scale of 0.85 arcsec per pixel on its CMOS cameras. Each exposure was 600 s in length. The “reach” of an observation obtained by Condor observations is the product of the total objective area and the total exposure time devoted to the observation. As Condor consists of six individual telescopes, each of objective area 0.0254 $m^{2}$, a one-second exposure with one telescope of the array yields a reach of 0.0254 $m^{2}$ s, and a one-second exposure with the entire array (i.e. with all six 
telescopes) yields a reach of 6 x 0.0254 $m^{2}$ s = 0.153 $m^{2}$ s. The dates of observation and reaches of the images of T CrB are presented in Table 1. Three or four telescopes were often used simultaneously with H$\alpha$ filters, accounting for the deeper reach in that filter.

The observation log, filter central wavelengths, and reach are summarized in Table 1. The array was dithered by a random offset of 15 arcmin between exposures. Images of the dusk and dawn twilight sky were obtained every night, and bias observations were obtained at the end of every night.

The observations were processed through the Condor data pipeline, with steps involving bias subtraction, flat fielding and background subtraction, astrometric calibration, and photometric calibration. As described in \citet{Lanzetta2023}, the astrometric calibration yields systematic uncertainties of $\lesssim$ 0.1 arcsec.

\begin{table}[ht]
\centering
\begin{tabular}{p{1.00in}cccc}
\multicolumn{4}{c}{{\bf Table 1:}  Details of Imaging Observations} \\

\multicolumn{1}{c}{} &\multicolumn{1}{c}{J2000} & \multicolumn{1}{c}{R.A. = 17:50:13} &\multicolumn{1}{c}{Dec = $-$06:42:28}\\
\hline
\hline

\multicolumn{1}{l}{Filter} & \multicolumn{1}{c}{Start Date} & \multicolumn{1}{c}{End Date} &\multicolumn{1}{c}{Reach ($m^{2}$s)} \\
\hline
\multicolumn{1}{l}{Luminance 550.0 nm}  & 2024-05-28 & 2024-05-30 & 366.4 \\
\multicolumn{1}{l}{[HeII] 468.6 nm}    & 2023-06-05 & 2023-06-08 & 1,511.5 \\
\multicolumn{1}{l}{[O III] 500.7 nm}    & 2023-06-05 & 2024-05-29 & 1,939 \\
\multicolumn{1}{l}{[HeI] 587.6 nm}    & 2023-06-06 & 2024-06-09 & 1,007.1 \\
\multicolumn{1}{l}{H$\alpha$ 656.3 nm}   & 2023-08-05 & 2024-05-30 & 9,557.8  \\
\multicolumn{1}{l} {[N II] 658.4 nm}     & 2024-06-04 & 2024-05-30 & 3,313.2 \\
\multicolumn{1}{l} {[S II] 671.6 nm}     & 2023-06-05 & 2024-05-30 & 3,511.7 \\

\hline

\end{tabular}
\end{table}

\newpage
\subsection{SALT spectroscopy}
 Three spectra of the RS Oph ejecta detected by Condor, each one hour in length, were obtained with the Southern African Large Telescope (SALT) and its Robert Stobie Spectrograph \citep{odon2006,Kobulnicky2003} using the PG2300 grating on 17 July, 21 July and 10 August of 2023. Position angles of 270 (East-West orientation), and 180 and 180 (North-South orientation) degrees were used, respectively, on those nights. 
 
 The grating/spectrograph covered the wavelength range 593.3 to 677.6 nm at a dispersion of 0.013 nm per unbinned pixel, with a resolution of 0.22 $\pm$ 0.025 nm. The data were reduced with the SALT RSS pipeline described in \citet{Kniazev2022}, which includes correction of bad columns and lines on the CCDs; calculation and application of gain-correction coefficients; construction of a spectral flat field for subsequent pixel sensitivity correction; and location and removal of cosmic ray events. These steps were followed by reduction of a ThAr reference spectrum, reduction of a spectrophotometric standard (LTT 4816), and reduction of the object. 

\section{The extended nebulosity of RS Oph}

\subsection{Narrowband Images}
The six Condor narrowband images of RS Oph (HeII, [OIII], HeI, H$\,\alpha$, [NII] and [SII]) are shown as a mosaic in Figure \,1, with RS Oph indicated by pairs of red tick marks. The coherent, shell-like structure surrounding RS Oph is most apparent in H$\,\alpha$, especially when contrasted with the images taken through the other filters. Similar nebular morphology is weakly discernible in [NII] and [SII],  and entirely absent in HeI, HeII and [OIII]. A linear feature at the southern boundary of the nebulosity (running NE to SW) is prominent in Figure 1.

%The six difference images (narrowband minus Luminance), are shown in Figure 2. 
The six difference images are shown in Figure 2. Each difference image was created by subtracting the Luminance image from the narrowband image on a pixel-by-pixel basis. The background was subtracted from each image before differencing, without further scaling. It is again seen that the $\sim$ 1.5 deg diameter structure surrounding RS Oph is most apparent in H$\alpha$.  An outline of the brightest H$\alpha$ nebular regions, suggestive of two lobes emanating from RS Oph, is shown in Figure 3. 

In Figure 4 we show a color composite which is a sum of the three difference images of the H$\alpha$, [NII], and [SII] minus Luminance, respectively. The structures seen in Figure 2 are more evident. In Figure 5 we show the ``star-cleaned" version of Figure 4, created with the V2 version of StarNet2 (\url{https://www.starnetastro.com/experimental}).

The brightest areas, comprising $\sim$ 1\% of the area in the H$\alpha$ image are radiating with a surface brightness $\sim$ 1.4 x $10^{-17}$\, $\mathrm{erg\, sec^{-1}\, cm^{-2}\, arcsec^{-2}}$.  The monochromatic 1-sigma uncertainty per pixel of the H$\alpha$ image is $\sim$ 0.5 microJy, which corresponds to an energy flux 1-sigma uncertainty per pixel of 3.5 $\times$ $10^{-18}$ $\mathrm{erg\,sec^{-1}\,cm^{-2}}$; a surface brightness 1-sigma uncertainty over one pixel of 4.8 $\times$ $10^{-18}$\, $\mathrm{erg\,s^{-1}\,cm^{-2}}$; and a surface brightness 3-sigma uncertainty over a 10 $\times$ 10 $\mathrm{arcsec^{2}}$ region of 1.2 x $10^{-18}$\, $\mathrm{erg\,s^{-1}\,cm^{-2}}$. 

\begin{figure*}[h!]
\vspace{0.5 cm}
\includegraphics[width=0.9\textwidth]{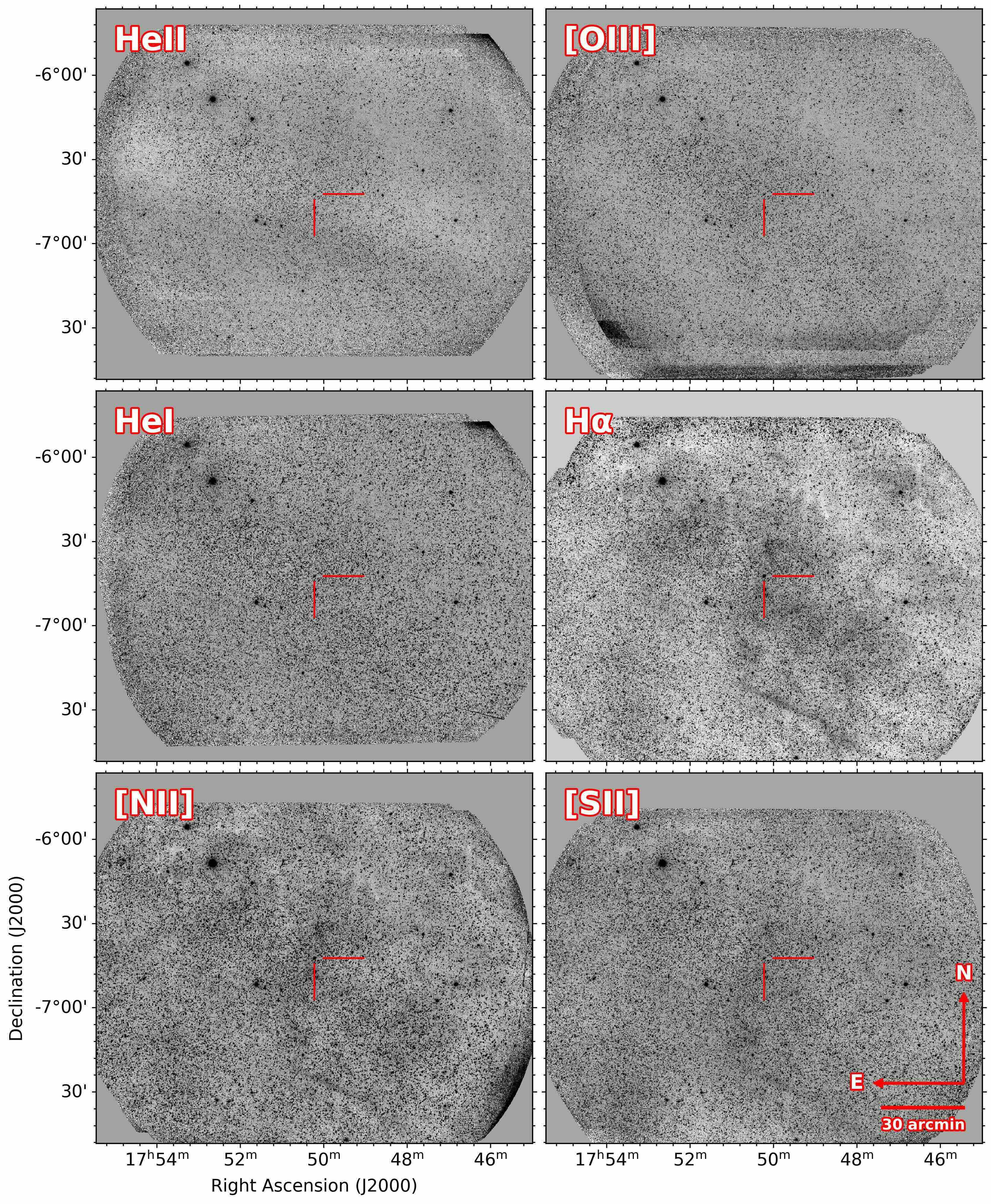}
\caption{Condor narrowband images of the area surrounding RS Oph in HeII, [OIII], HeI, H$\,\alpha$, [NII]] and [SII] filters. RS Oph is marked with red ticks. Images have been 8x8 block-summed and are displayed with linear scaling. The RS Oph nova super-remnant is immediately apparent only in the H$\,\alpha$ image as a nebulosity filling much of the field-of-view, and bounded by a nearly-one-degree long linear feature running NE to SW at the southern edge of the image. }
\label{fig:six_panel}
\end{figure*}

\newpage 
\begin{figure*}[h!]
    \includegraphics[width=0.9\textwidth]{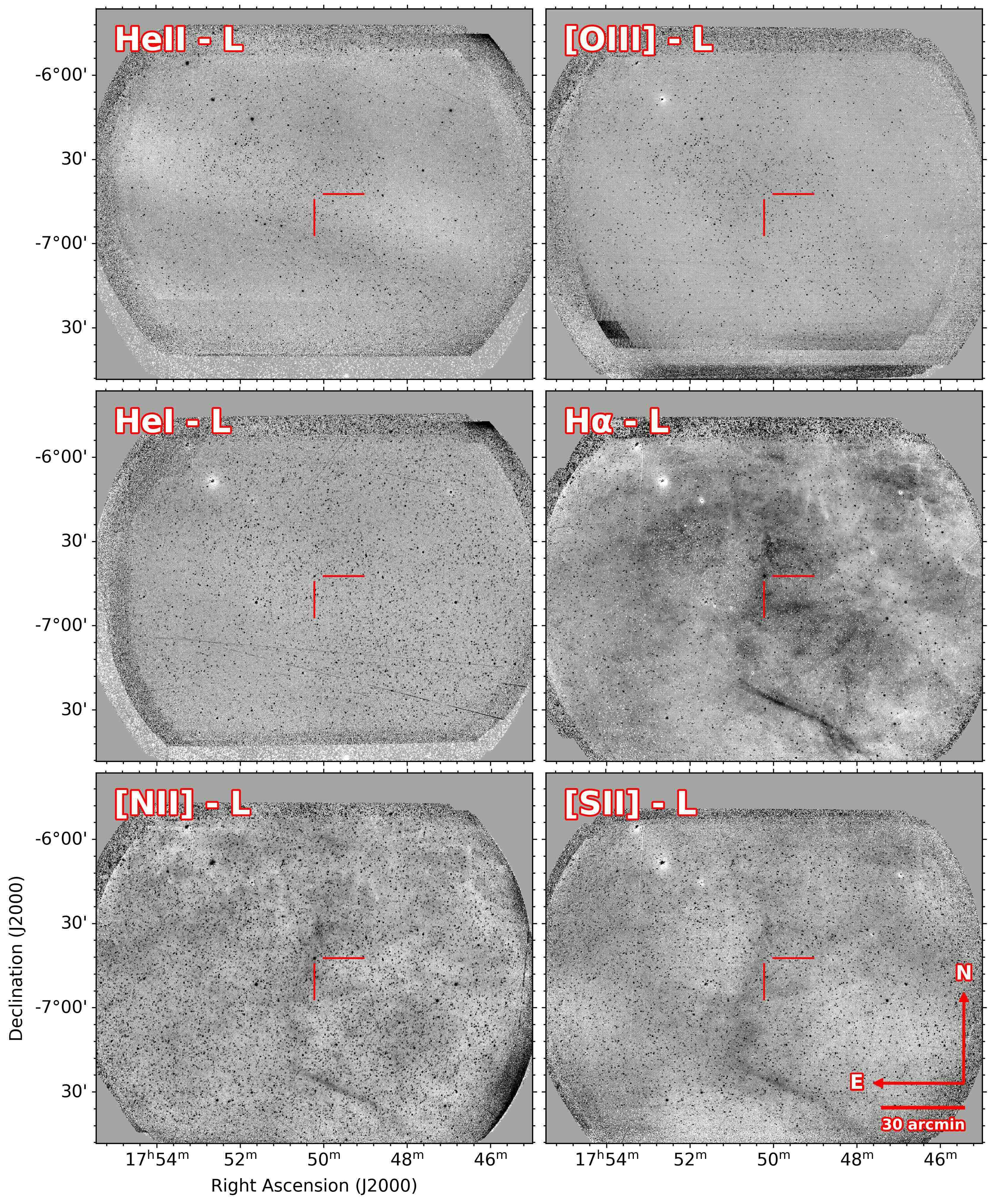}
    \caption{Difference images (narrowband minus Luminance) of the area surrounding RS Oph. Images have been 8x8 block-summed and are displayed with linear scaling. RS Oph is marked with red ticks. The nova super-remnant is now apparent in the H$\,\alpha$, [NII]] and [SII] filters.}
   
\end{figure*}

\newpage

\begin{figure*}[h!]
\vspace{0.5 cm}
    \includegraphics[width=0.9\textwidth]{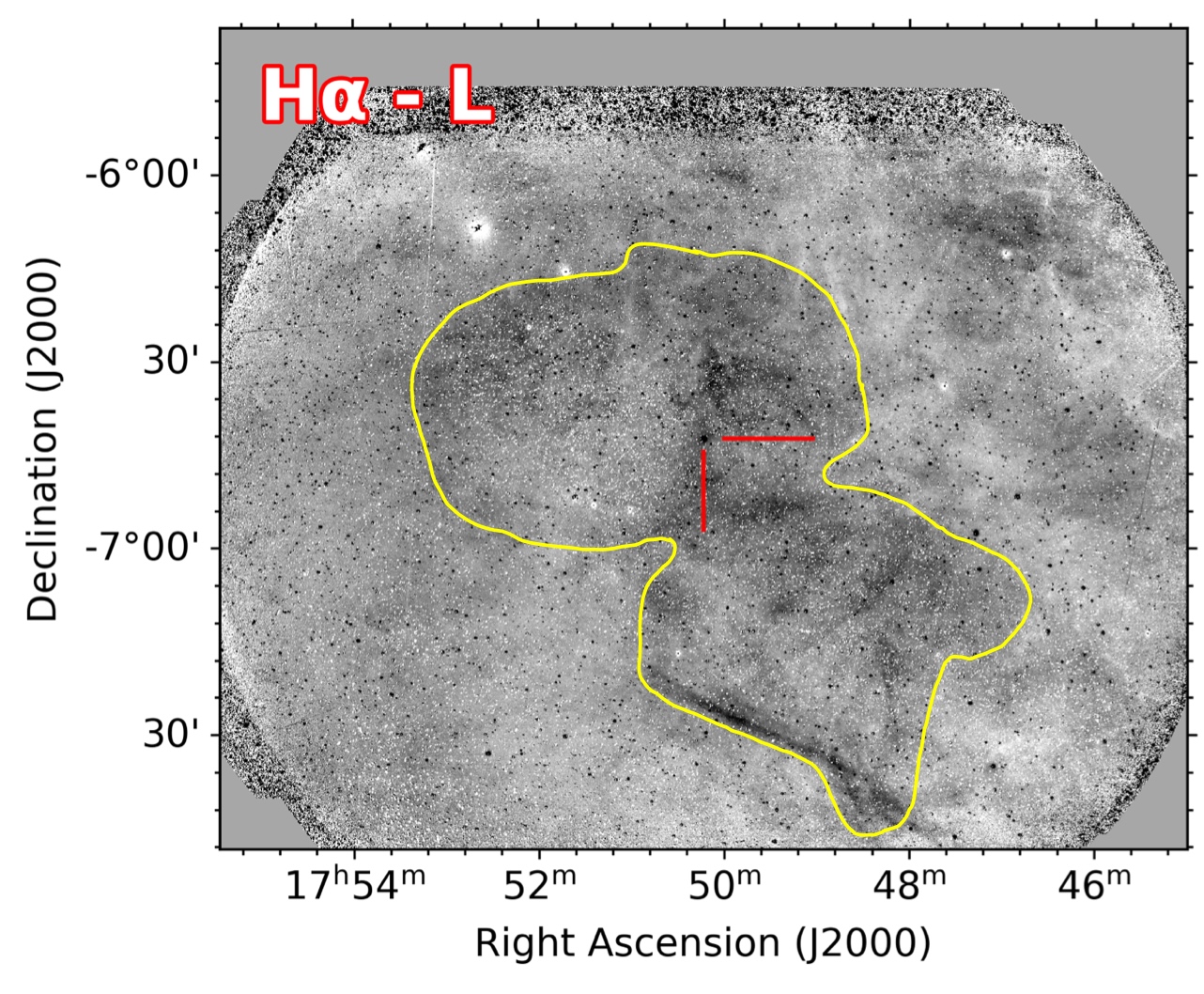}
    \caption{H$\alpha$ - Luminance difference image of RS Oph NSR (same as in Figure 2) with an overlaid sketch of the most prominent H$\alpha$ emission. Two lobes of nebulosity are seen.}
   
\end{figure*}

\newpage

\begin{figure*}[h!]
    \includegraphics[width=0.9\textwidth]{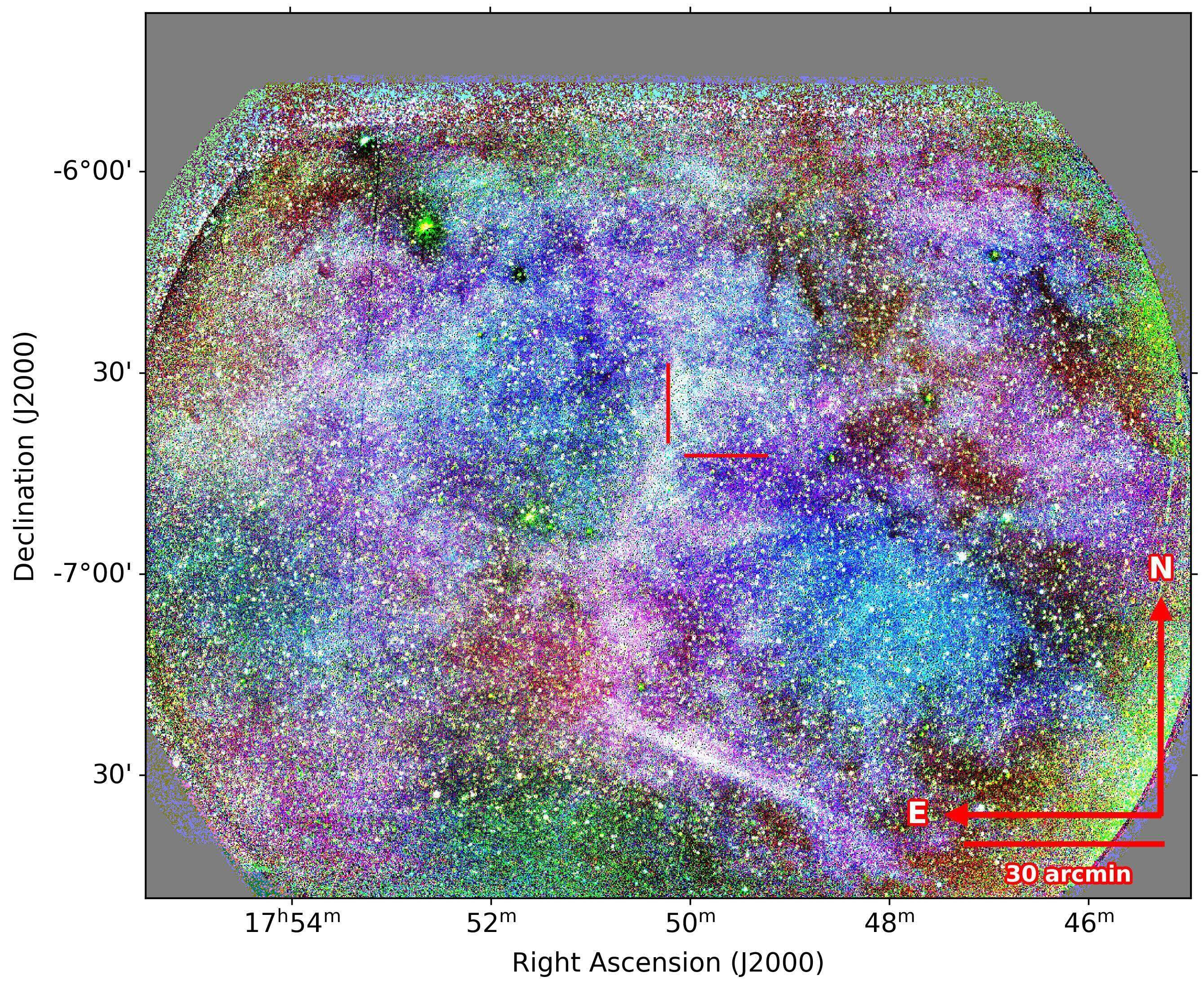}
    \caption{RGB image of RS Oph (position indicated with red ticks) and its surrounding emission. Each of the three channels is a block-summed difference image (see Figure 3). Blue corresponds to H$\alpha$ - Luminance; green corresponds to [NII] - Luminance; and red corresponds to [SII] - Luminance. The intensity scaling of each channel has been adjusted individually to highlight the emission features.}
\end{figure*}

\newpage

\begin{figure*}[h!]
    \includegraphics[width=0.9\textwidth]{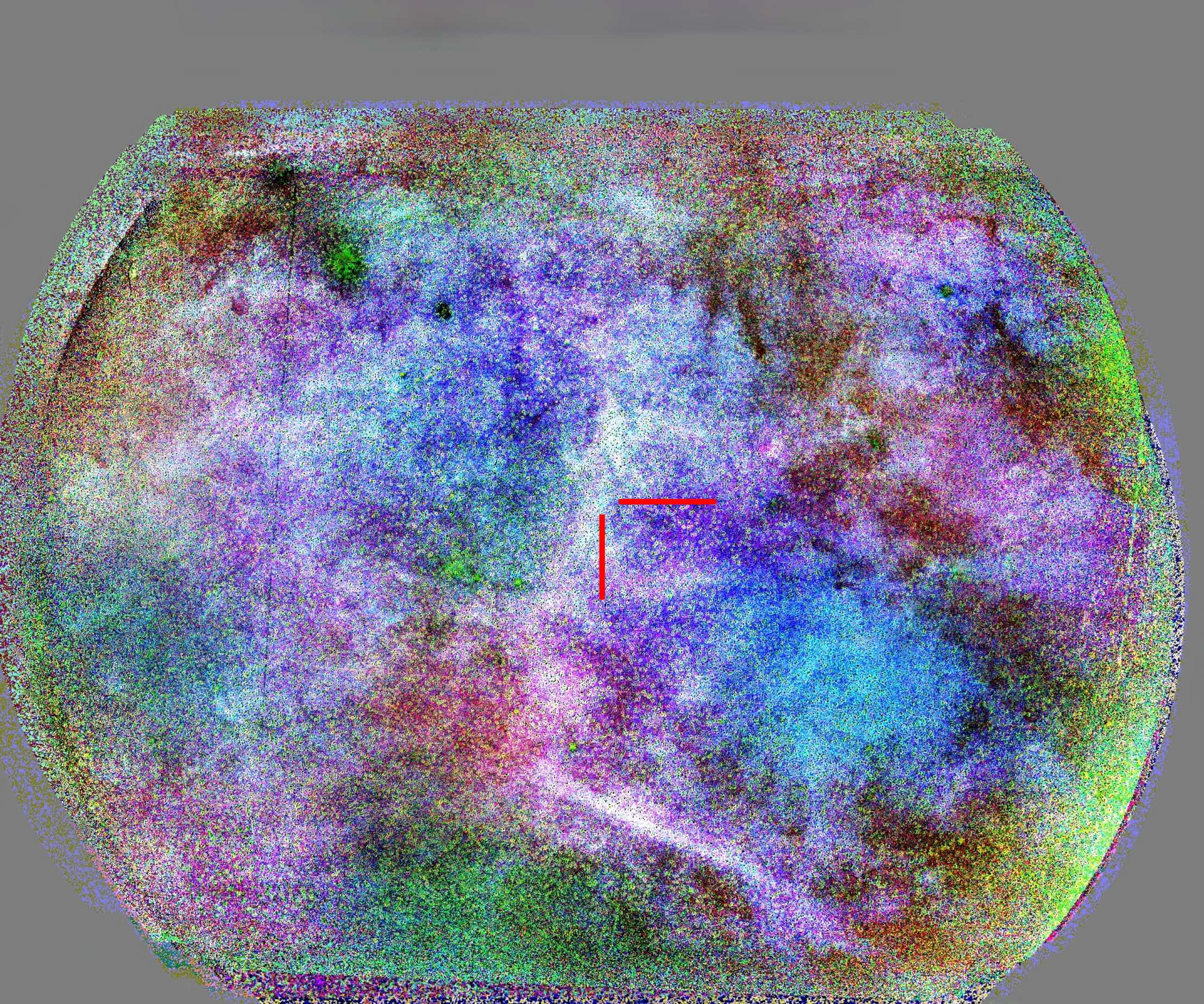}
    \caption{Same as Figure 4, but after the ``star-cleaning" algorithm STARNET2 is applied. The location of RS Oph is indicated with red ticks.}
    \label{fig:six_panel_sub}
\end{figure*}

\newpage

\section{Spectra of the RS Oph NSR}
\subsection{Slit locations}

As noted in Section 2, three spectra (each of one hour duration) of the RS Oph nebula were taken with SALT. The three slit placements are shown in Figure\, 6. Spectra were taken with the yellow-colored slit position on 17 July 2023, the cyan-colored slit position on 21 July 2023, and the magenta-colored slit position on 10 August, 2023. RS Oph is indicated by a pair of diagonal red lines. The white arrow in Figure 6 indicates the direction of proper motion and the angular distance that has been traversed by RS Oph in the past 50 kyr.

\begin{figure*}[h!]
    \includegraphics[width=0.95\textwidth]{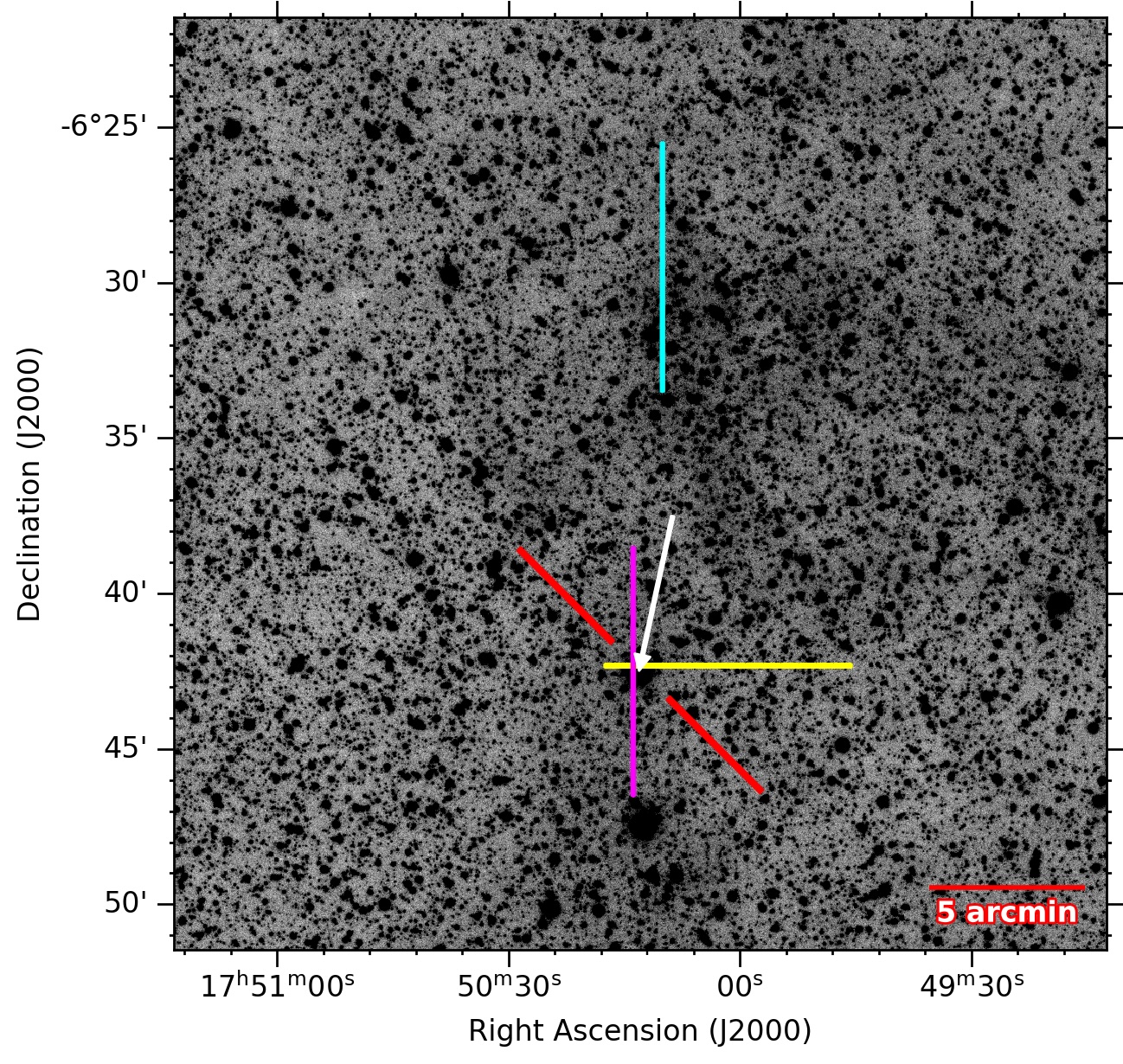}
    \caption{H$\alpha$ image of the region surrounding RS Oph, which is indicated by diagonal red ticks. The proper motion of RS Oph (with a Gaia-measured amplitude of 6.11 mas/yr) is shown as a white arrow, whose length corresponds to the motion of RS Oph in the past 50 kyr. The positions of two inner and one outer SALT RSS slits, described in section 2, are shown as yellow, magenta and cyan lines, respectively. The image has been smoothed via a Gaussian kernel ($\sigma = 3$ pixels) and is displayed with linear scaling. The direction of proper motion of RS Oph is approximately perpendicular to the linear feature marked by yellow brackets in Figure 1, suggesting that this feature might be material snow-ploughed by RS Oph.}
    \label{fig:Ha_all_slits}
\end{figure*}

\newpage

\subsection{Spectra}

The reduced SALT 2-D spectra of the inner nebulosity of RS Oph, taken on 10 August 2023 (with the magenta-colored slit in Figure 6), are shown in Figures 7 and 8. (The other inner slit spectrum was overwhelmed by light from RS Oph itself, and is unusable). The faint, curved features between $\sim$ 663.0 - 666.5 nm in the the 2D spectra are residual artifacts due to slightly imperfect correction of the gain of the two amplifiers on each of the RSS CCD detectors. 

To subtract the night sky lines, we used the ‘background’ task from the ‘longslit’ package in IRAF with parameters described in detail in \citet{Kniazev2022}. Specifically, for each selected wavelength (column in the two-dimensional spectrum), the night sky background was approximated by a third-degree polynomial, and all points exceeding two sigma errors were removed from the approximation, assuming that they belonged to the object
spectrum rather than the night sky background. This procedure was performed iteratively ten times for each wavelength of the two-dimensional spectrum, and the result of this iterative approximation was considered to be the found spectrum of the night sky, which was subsequently subtracted. A closeup of the region around H$\alpha$ (the bottom of Figure 8) shows no trace of e.g. the bright OH 6577 night-sky line.

Within $\sim$ 40" of the star the only emission line clearly detected, after night-sky subtraction, is the H$\alpha$ line. It is split into two components (see the sky-subtracted spectrum of Figure 8) with radial velocities of $\sim$ $\pm$ 60 km/s relative to that of RS Oph itself. The wings of the H$\alpha$ line of the clearly detected, extended nebulosity centered on RS Oph extend out to $\sim$ +/- 35 arcsec from RS Oph itself.

Numerical models \citep{Healy-Kalesh2023} demonstrate that thousands of closely spaced (in time) RN eruptions are able to clear large cavities around their host stars. The nova ejecta and the swept-up ISM create dynamic NSRs tens of parsecs in radius. These NSRs inevitably contain a low-density cavity, bordered by a hot ejecta pile-up region, which is surrounded by a cool, high-density, thin shell expanding at tens of km/s. The small expansion velocity is because the swept-up matter is hundreds to thousands of times more massive than the aggregate sum of all the matter ejected by the WD. This is in sharp contrast to the observed shells of non-recurrent novae, which display expansion velocities of hundreds to thousands of km/s. The observed $\sim$ $\pm$ 60 km/s H$\alpha$ line matches the \citep{Healy-Kalesh2023} prediction of the expected expansion velocity of a NSR seen near its center (i.e. close to the line-of-sight to the nova itself). 

The H$\,\alpha$ line radial velocities of the outer shell (see Figures 9, 10 and 11), over the $\sim$ 400 arcsec length of the RSS slit, are all $\sim$ -14 km/s, which is close to the RS Oph systemic velocity value of -38.7 $\pm$ 0.4 km/s \citep{Brandi2009}.  The $\sim$ 25 km/s difference between these velocities is likely due to the deceleration of the RN ejecta, whose velocity has been modified by the copious ISM mass that it has swept up.

The complete lack of higher velocities (multiple hundreds or thousands of km/s) argues against the nebulosity of Figures 1 and 2 being due to a supernova remnant despite the presence of [SII] lines. As already noted, the lack of [OIII] emission in the images of Figure 1 provide further evidence against a supernova remnant interpretation, and simultaneously rules out a planetary nebula. The logarithms of the ratios of the [NII]658.4 /H$\alpha$ and [SII]/H$\alpha$ lines are both $\sim$ -0.45 over the length of the slit (see Figure 11). Both values are large enough to be indicative of shock ionization \citep{Kniazev2008}, as expected for an NSR.

The outer shell spectrum (Figures 9, 10, 11 and 12) shows both components of the density-sensitive [SII] 671.6/673.1 doublet. The line ratio's value of 1.41 $\pm$ 0.05 (see Figure 12) is indistinguishable from the 1.45 value of the low density limit $\sim$ 10 electrons/$\mathrm{cm^{3}}$ (cf. Figure 2 of \citet{LeTiran2011}). We use this value to roughly estimate the shell mass below. 

\newpage 
\begin{figure*}[h!]
\includegraphics[width=0.7\textwidth]{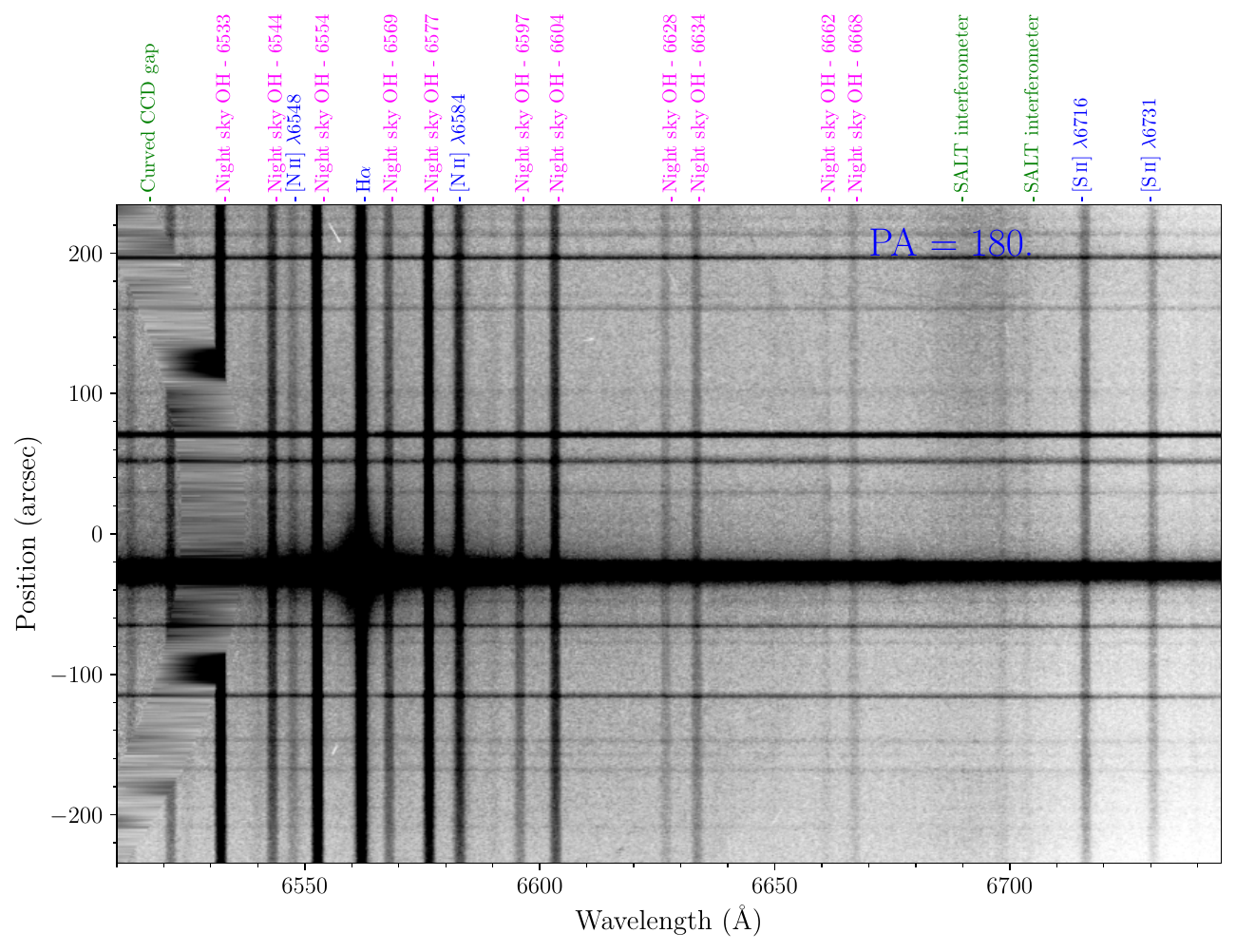}
\caption{SALT spectrum of the inner RS Oph shell taken on 10 August 2023 with the magenta colored slit of Figure 6. All unidentified (vertical) lines are due to the night sky. Unidentified horizontal lines are field stars that were aligned with the slit during the observations. }
\label{fig:PG2300_inside}
\end{figure*}
%\captionsetup{justification = raggedright, singlelinecheck = false}

\newpage
\begin{figure*}[h!]
\includegraphics[width=0.7\textwidth]{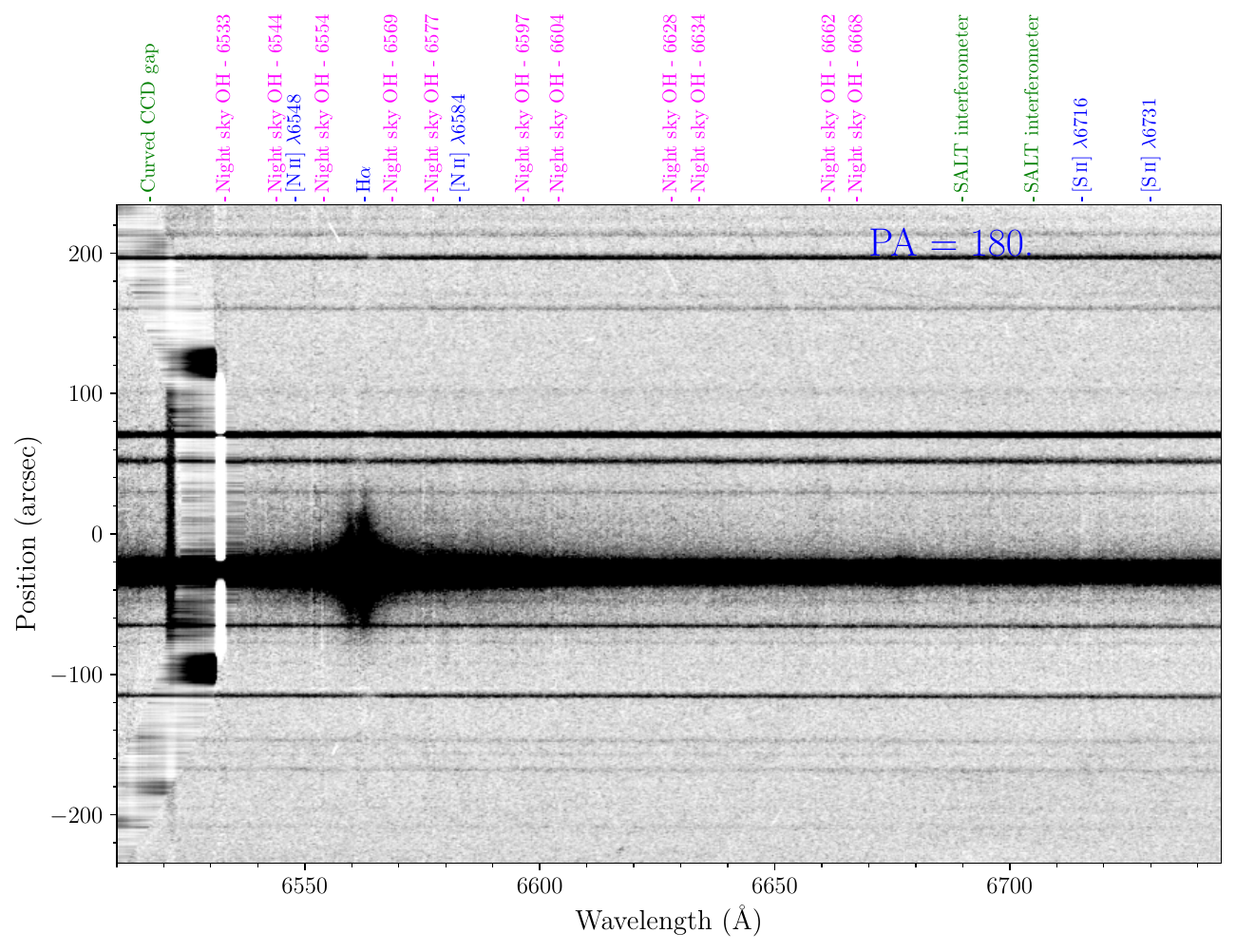}
\label{fig:PG2300_inside}
\hspace{-12.8 cm}
\raisebox{-8cm}{\includegraphics[width=0.7\textwidth]{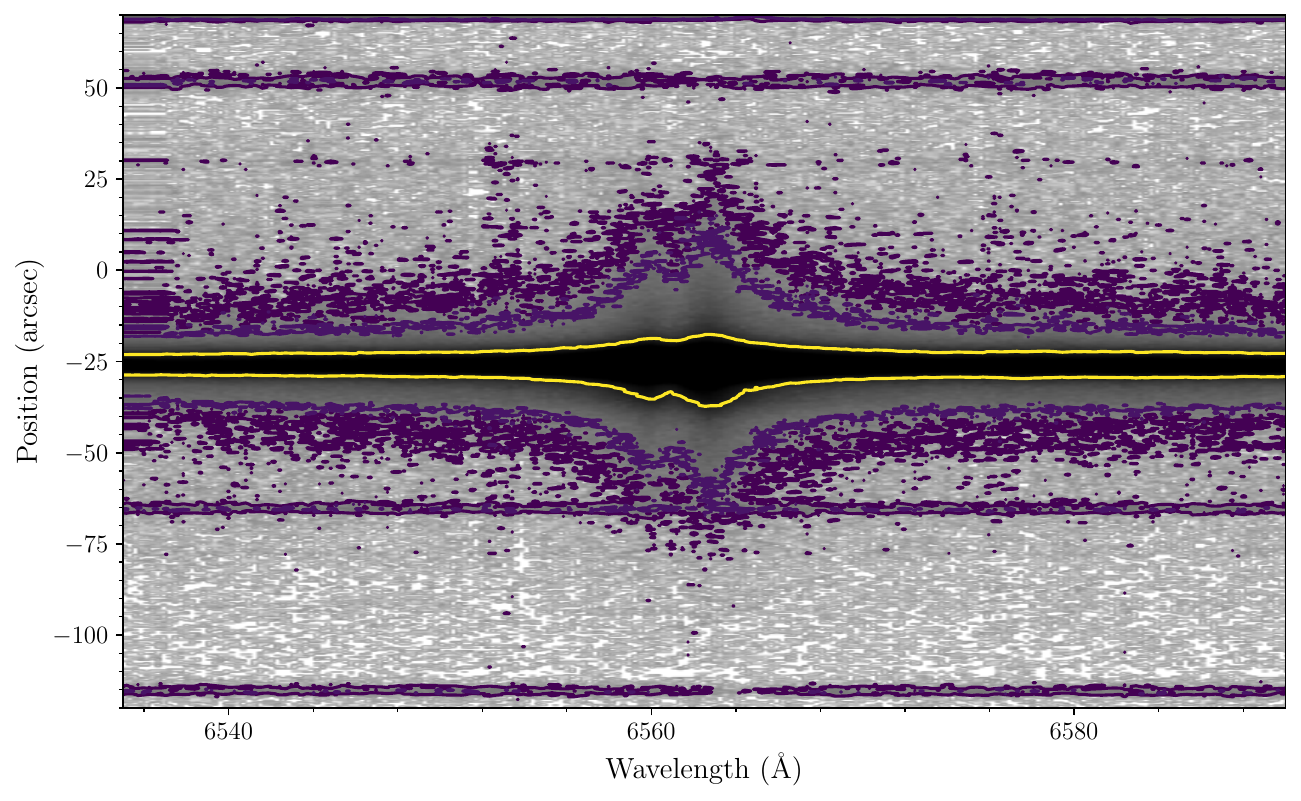}}
\caption{Top: Same as Figure 7, but the spectrum is sky-subtracted. Bottom: A closeup of the sky-subtracted region around H$\alpha$ in Figure 8a. The black, grey and purple pixels are, respectively, at least 200, 20 and 2 counts above background.}
\end{figure*}

\newpage
\begin{figure*}[h!]
    \includegraphics[width=0.7\textwidth]{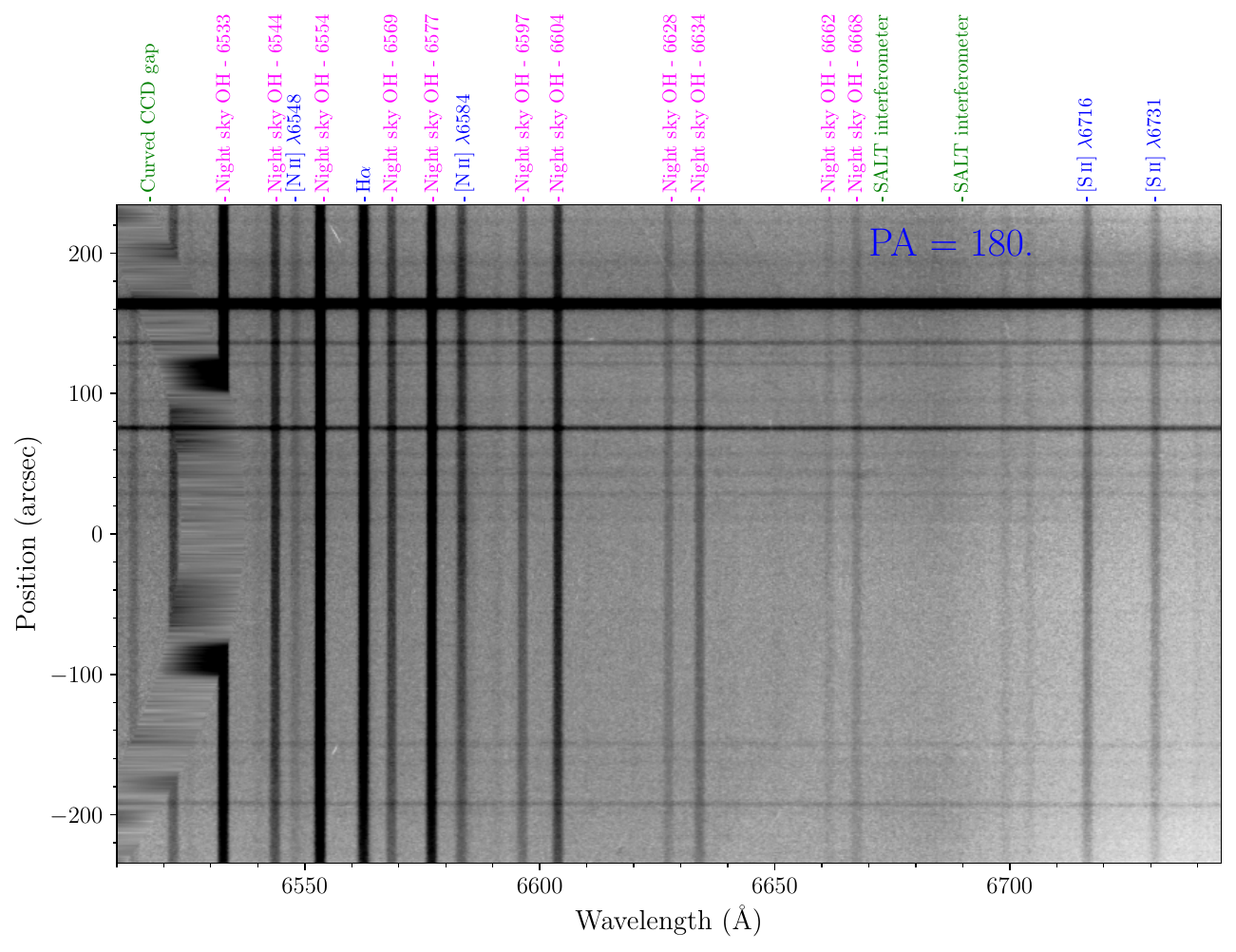}
    \caption{SALT spectrum of the outer RS shell shell, taken with the cyan-colored slit of Figure 6. All faint unidentified lines are due to the night sky.}
    \label{fig:PG2300_outside}
\end{figure*}

\begin{figure*}[h!]
    \includegraphics[width=0.70\textwidth]{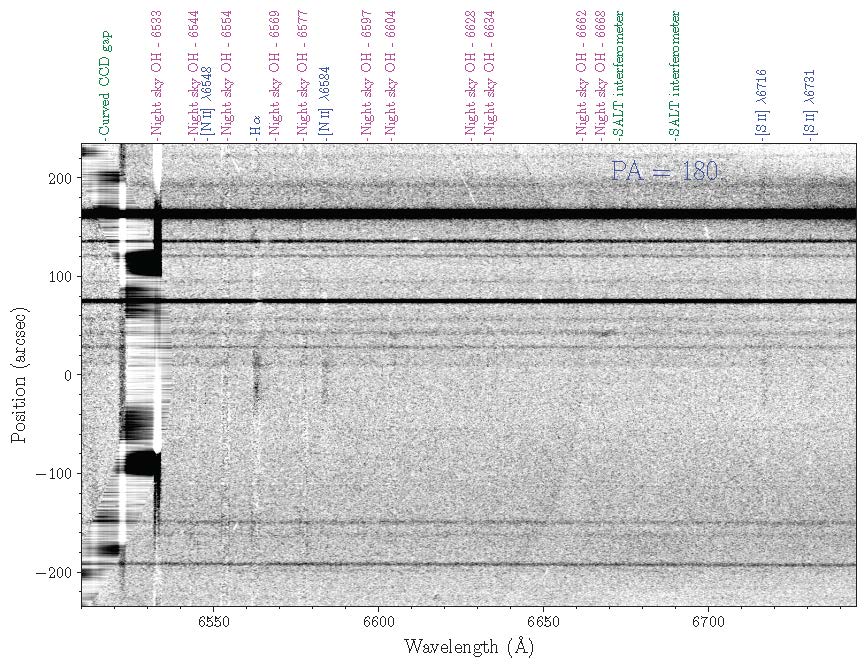}
    \caption{Same as Figure 9 but the spectrum is sky-subtracted. All faint unidentified (vertical) lines are due to imperfectly subtracted night sky.}
    \label{fig:PG2300_outside}
\end{figure*}

\newpage

\begin{figure*}[h!]
    \includegraphics[width=0.39\textwidth]{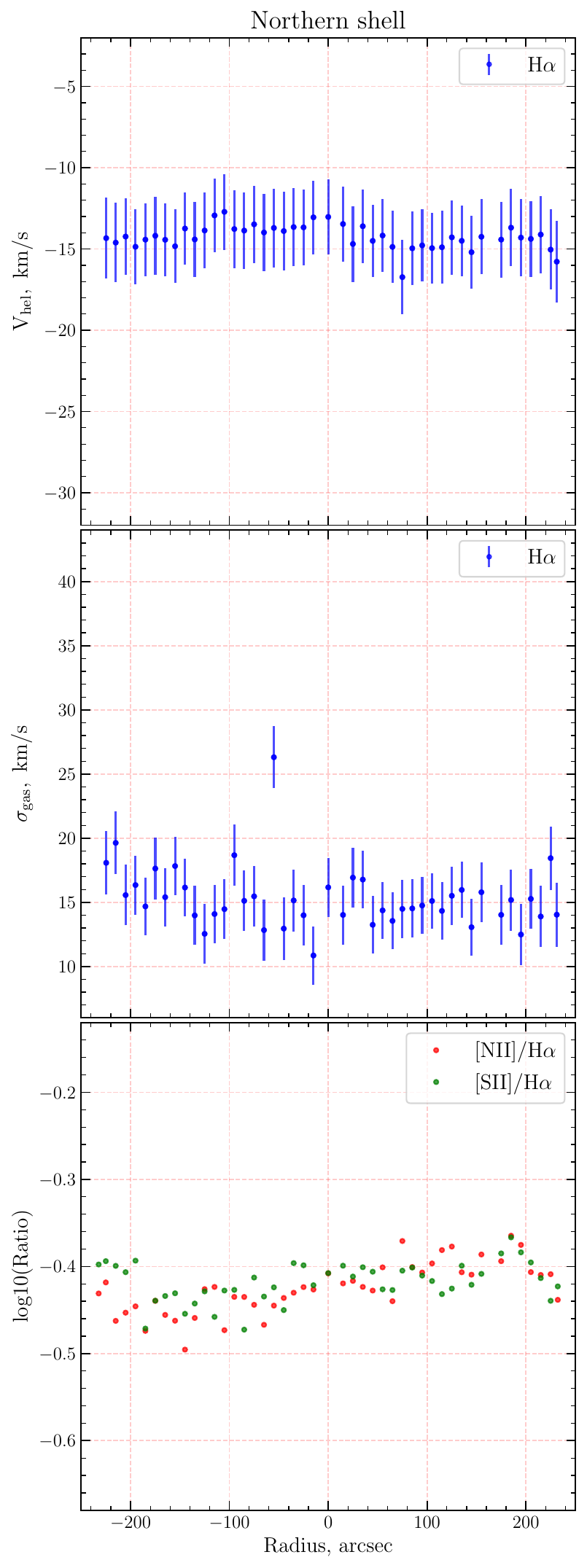}
    \caption{Top: Measured radial velocity of the H$\alpha$ line along the extent of the Northern (cyan-colored) slit. Data were binned into 10 arcsec-wide bins to reduce noise. The velocity of each binned data point was determined relative to the nearest night sky line at 655.35515 nm. The average velocity of H$\alpha$ along the slit is -14.25 $\pm  0.74$ km/s. Middle: The FWHM of the H$\alpha$ line along the Northern slit is $\sim$ 15 km/s, corresponding to 2.5 pixels on the detector. This demonstrates that the H$\alpha$ line is unresolved. Bottom: The logarithms of the ratios of the [NII]658.4 /H$\alpha$ and [SII]/H$\alpha$ lines, both $\sim$ -0.45, indicative of shock ionization. See text.  } 

\end{figure*}

\newpage

\begin{figure*}[h!]
    \includegraphics[width=0.9\textwidth]{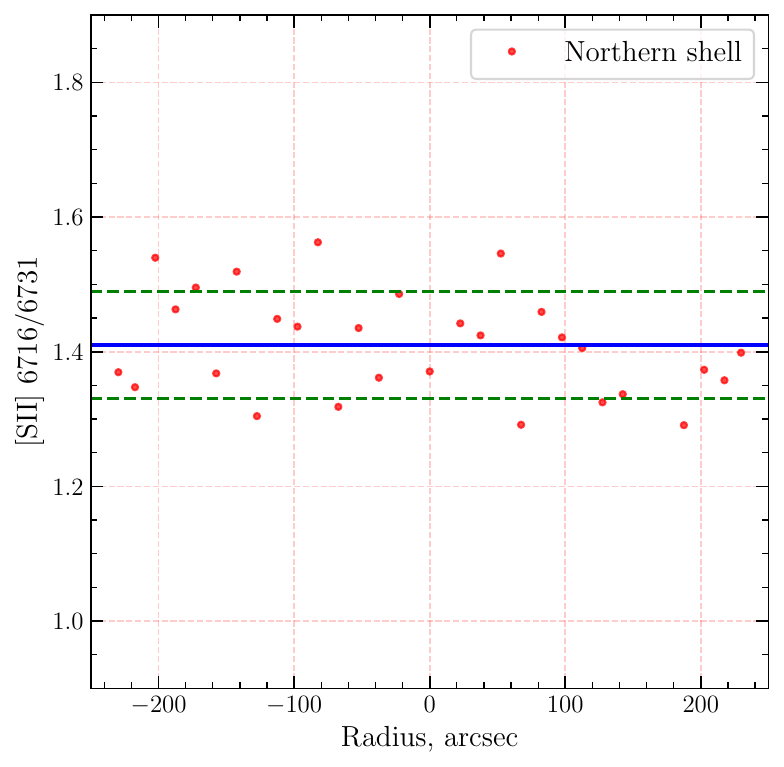}
    \caption{The ratio of the forbidden [SII] line doublet 6716/6731 along the cyan-colored, ``outer" shell SALT slit seen in Figure\,6. The value of the line ratio is $\sim$ 1.41 $\pm$ 0.05, with individual points' errors being 
    $\sim \pm 0.10$ consistent with an electron density of $\leq$ $\sim$ 50 electrons/$\mathrm{cm^{3}}$.} 

\end{figure*}

\newpage
\section{The Nova Super-remnant}

\subsection{Motion of RS Oph and the shell's age}

The $\sim$ 1.5 degree angular size of the RS Oph nebulosity seen in Figure \,2, at the 2.68 $\pm$ 0.16 kpc \citep{schaefer2022} distance of RS Oph, corresponds to $\sim$ 70 pc. This is $\sim$ half the size of the 134 pc super-remnant of M31-12a, and somewhat larger than the NSRs of KT Eri ($\sim$ 50 pc) and T CrB ($\sim$ 30 pc). It is two orders of magnitude larger than the $\sim$ 0.1 - 0.5 pc-sized ejecta of 19th and 20th century novae. 

RS Oph last erupted as a nova in 2021, and the maximum speed of its ejecta was $\simeq$ 3600-5600\,km\,s$^{-1}$ \citep{Bode2007,tomov2023}. Even ignoring the deceleration as those ejecta interact with the surrounding interstellar medium \citep{Duerbeck1987,santa2020}, it would take the ejecta $\sim$ 10 kyr to achieve the observed size of the RS Oph shell. In fact, nova ejecta sweep up of order their own mass on timescales of order 1000 yr \citep{santa2020}, so that the nebulosity that we detect must be multiple tens of thousands of years old, produced by thousands of RS Oph nova eruptions. RS Oph's presence within its own shell (Figure\,2) is consistent with the shell's ejection history.

\subsection{Shell mass}

Hydrodynamic 1D simulations of many thousands of RN ejection events \citep{Healy-Kalesh2023} leading to a NSR suggest that their boundary thicknesses are typically $\sim$ 10\% of the radius of the NSR, with the vast majority of mass being swept-up ISM. There is an indication of a sharp boundary in the form of an H$\alpha$ linear feature $\sim$ 45' to the south and SW of RS Oph (see Figure 2, 3, 4 and 5) indicative of a shell edge. However the rest of the NSR displays a patchy morphology, which is not surprising if the density of interstellar medium (ISM) surrounding RS Oph was itself irregular.

The simplest model possible of RS Oph's NSR is a sphere of radius 35 pc whose boundary is 3.5 pc thick, encompassing swept-up ISM and nova ejecta with a uniform density of $\sim$ 5-50 H atoms/$\mathrm{cm^{3}}$ (assuming, of course, that the [SII] lines of Figures 8, 9, 10 and 11 arise in gas associated with the NSR). The mass enclosed by the shell's boundary layer would then be $\sim$ 20 - 200 $M_{\odot}$. A significantly more accurate mapping of the super-remnant density will be needed to improve on this very rough estimate. But a simple order-of-magnitude ``sanity check" of this mass estimate is straightforward, as follows.

The RS Oph WD must be accreting mass from the red giant wind at about $10^{-7}$$M_{\odot}$/yr \citep{Booth2016}, so that the total mass ejected per outburst is $\sim$ 2 x $10^{-6}$$M_{\odot}$. The observed ejection velocities, corrected for inclination are $\sim$ 5000 km/s \citep{Bode2007}, so that the kinetic energy carried away by one eruption is $\sim$ 
5 x $10^{44}$ erg. Of order 2000 eruptions, spaced over 40,000 yr, have injected $10^{48}$ erg into the swept-up ISM. Assuming that half that energy has been radiated away, and half is retained as the kinetic energy of the NSR expanding at $\sim$ 50 km/s, we derive an NSR mass of $\sim$ 20 $M_{\odot}$, consistent with our earlier rough estimate.

\section{Conclusions}

Deep narrowband Condor Array Telescope imagery of the environs of the recurrent nova RS Oph have revealed a $\sim$ 1.5 degree, $\sim$ 70 pc-sized, possibly bi-lobed nebulosity surrounding the star: a nova super-remnant. Such nebulae are {\it predicted} to be intrinsic and unavoidable features surrounding {\it all} RNe \citep{Healy-Kalesh2023}, and its detection here strongly supports the theoretical model of the pile-up of thousands of ejected shells. The RS Oph super-remnant must be at least tens of thousands of years old to be consistent with RS Oph's (Gaia-determined) proper motion and the $\sim$ 3600 km/s initial speed of ejection of its nova events. The lack of significant [OIII] emission rules out a supernova or planetary nebula origin for the observed nebulosity. The logarithmic ratios of the [NII]658.4/H$\alpha$ and [SII]/H$\alpha$ lines, both $\sim$ -0.45, are indicative of shock ionization as expected for a NSR. The observed line ratio [SII] 6716/6731 is consistent with a matter density of $\leq$ 50 atoms/$\mathrm{cm^{3}}$. A simple model (Section 5.2) of a (mostly hollow) NSR shell then yields a shell mass of $\sim$ 20 - 200 $M_{\odot}$, expanding at a few tens of km/s, with an age of order 50-100 kyr.

\section{Acknowledgements}
MMS and JTG acknowledge the support of NSF award 2108234. KML is supported by the National Science Foundation under grants 1910001, 2107954, and 2108234. AK  acknowledges the Ministry of Science and Higher Education of the Russian Federation grant 075-15-2022-262 (13.MNPMU.21.0003). JM acknowledges support from the Polish National Science Center grant 2023/48/Q/ST9/00138. This work has made use of data from the European Space Agency (ESA) mission {\it Gaia} (\url{https://www.cosmos.esa.int/gaia}), processed by the {\it Gaia} Data Processing and Analysis Consortium (DPAC, \url{https://www.cosmos.esa.int/web/gaia/dpac/consortium}). Funding for the DPAC has been provided by national institutions, in particular the institutions participating in the {\it Gaia} Multilateral Agreement.

{}
\bibliographystyle{aasjournal}

\end{document}